# Time-dependent resonant magneto-optical rotation

Dariusz Dziczek *

*Institute of Physics, Nicolaus Copernicus University, Grudziądzka 5/7, 87-100 Toruń, Poland*

Results of a fairly straightforward experiment on resonant magneto-optical rotation by rubidium-87 atoms revealed strong time-dependence of the polarization plane of light emerging from atomic vapors following a sudden irradiation with a laser beam. The rotation of the plane appears as a not direct consequence of the influence of the magnetic field on atoms. Reported measurements conducted using a vapor cell without any buffer gas or an anti-relaxation wall coating show that transmitted light has initially the same (linear) polarization as the incident one. Rotation of the polarization plane caused by an axial magnetic field develops in time scales similar to the pace of establishing the optical pumping/relaxation equilibrium in the atomic ensemble. The traditional passive Faraday rotation picture providing working description for the resonant magneto-optical effects in steady-state conditions does not explain the observed sequence of evolution of the polarization. The picture has to be augmented with analysis of time dependence in the spatial distribution of relevant quantum states within the atomic ensemble as well as longitudinal and transversal intensity profiles of the light beam. Isolation of external effects caused by the optical pumping and relaxation processes from consequences of internal atomic coherences should uncover more details of mechanisms of the latter in the presence of magnetic field. Plentitude of data available from time-dependent polarization measurements provides strong incentives for further development and tests of comprehensive models of coherent interactions of atomic ensembles with radiation and possibly a base for additional advance in understanding and applications of these phenomena.

* dziczek@fizyka.umk.pl

Coherent interactions of atoms with radiation have always been a very important part of experimental and theoretical atomic physics. Benefits from the development of laser technology in recent decades have facilitated unprecedented advance in this field. Significant contributions to the great progress are associated with tightly related studies of electromagnetically induced transparency (EIT) and resonant magneto-optical effects (MOEs). The wealth of work on the latter subject resulted in very impressive achievements in applications of atomic interactions for ultra-sensitive magnetometry [1,2].

The distinctive feature of resonant interactions is strong coupling of atoms with light. Mutual influence between atomic ensembles and radiation (especially the coherent one) leads to protuberant effects, which are relatively easy to notice and measure. Although far from being elusive, resonant phenomena are not, however, straightforward to model comprehensively. In the general view, the strong coupling lead to inhomogeneities in both sub-systems. Spatial light intensity distributions – already non-uniform in realistic situations – are altered by strong scattering, both longitudinally and transversally. Characteristics of atomic ensembles – populations of the relevant quantum states and coherences among them, also acquire significant spatial diversity. Even on this basis only, one can expect nonlinearities in observed dependencies of MOEs on the incident laser beam power. Understanding of the internal atomic factors responsible for nonlinearities requires determination of the external ones. Proper isolation of consequences of optical pumping in multi-level atomic energy structures from internal effects of light-induced coherences of quantum states, accounting also for atoms' motion and detuning from exact resonance is a challenging task. The contributors to its complexity have been incorporated in theoretical models successfully reproducing experimental results in a number of occasions [3], not delivering, however, any consistent general picture of it. Incidentally, premature conclusions have been drown based on not sufficiently comprehensive modelling.

Usual description of the resonant magneto-optical rotation (MOR) is based on the traditional pictures invented for the Faraday and Voigt effects in glasses and liquids which assume magnetically-induced birefringence of optical media [4]: Linearly polarized light is represented as two components of circular polarizations, their optical dispersion is affected by magnetic field in different ways so they acquire relative phase shift as they propagate. This results in rotated polarization plane when one combines the components switching back to the linear representation. This framework is filled up with appropriate physical mechanisms of influence of magnetic field on dispersion.

The above picture separates the action of magnetic field from the optical effect it induces, so it can be regarded as passive when considering the radiation. Obviously, it works well when the separation is the case, e.g. Faraday effect in flint glass. However, its limitations should be closely examined when light by itself affects media simultaneously with magnetic fields – as in all cases of resonant optical effects. Measurements reported in this work provide an example of situation in which the passive picture does not suffice for description of observations, creating needs for an extension or alternative approaches to the problem. Following considerations exploit an enforced role played by collective effects and the stimulated emission in a qualitative explanation of mechanisms of time-dependent resonant MOR in the case of moderate atomic concentrations. The same approach may lead to quite distinct views of the coherent interactions of atoms with radiation in general, and theoretical models constructed in ways emphasizing some aspects of resonant MOEs and other atomic coherence phenomena (like EIT) currently remaining unexplored.

Considering the rotation of polarization plane of resonant light emerging from a layer of atomic vapors placed in a homogeneous and constant magnetic field, following a sudden irradiation with a linearly polarized laser beam, one can find it difficult to formulate expectations based on information available from steady-state measurements. Transient problems are usually hard to solve and combined with complexities of interactions of realistic atoms with radiation comprise serious challenges. This motivated undertaking of the presented experimental work – in fact, to gather knowledge needed for more complete comprehension of details of experiments on EIT [5]. Qualitative analysis of the performed measurements has led to a conclusion that observations cannot be explained within the frame of the passive picture sketched above.

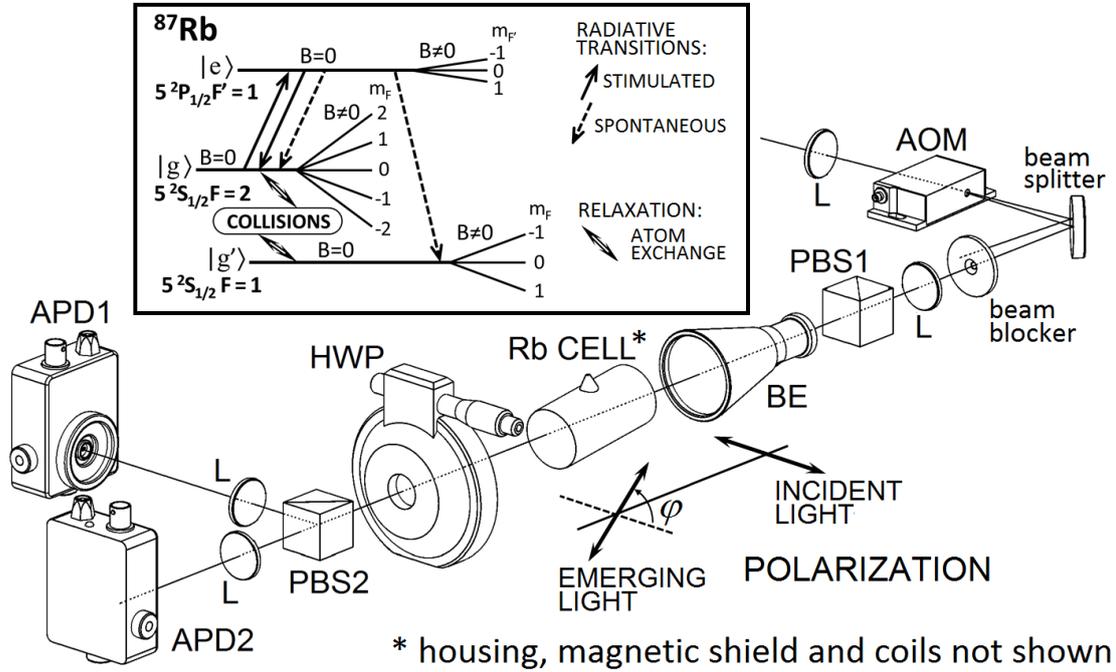

FIG. 1. Schematic of the apparatus for the measurements of time-dependent MOR and relevant $^{87}$Rb atoms' states with their Zeeman splitting; (L –lens, AOM –acousto-optic modulator, PBS – polarizing beam splitter, BE – beam expander, HWP – λ/2 plate, APD – amplified photodiode detector) .

The experiment had relatively simple concept and the apparatus was very similar to ones used in many steady-state studies. It is schematically depicted in Fig. 1. Target $^{87}$Rb atoms were confined in a cylindrical cell containing the natural mixture of isotopes and no gas admixtures. Concentration of atoms was controlled by precisely regulated temperature in special housing heated by circulation of hot air. A cylindrical, three-layer magnetic shield enclosed the housing to reduce influence of the ambient magnetic field. Two long coils inside the shield served to compensate the residual field in the housing and to apply precisely controlled axial test magnetic fields. A single laser beam was shaped with a polarization-maintaining optical fiber, lenses (L) and a beam expander (BE). An acousto-optic modulator (AOM) was used to control the intensity of light and a high-extinction-ratio polarizing beam splitter (PBS) cube assured its linear polarization. The incident beam was well collimated, had approximately Gaussian transversal intensity profile (only very slightly elliptical). The detector side of the setup consisted of a half-waveplate (HWP) in a precision rotation mount, another high-extinction-ratio PBS and two fast amplified photodiodes (APD). In the PBS-reflected beam's arm (vertical polarization), serving as the main detection channel, an additional Glan-Thomson polarizer (not included in Fig. 1) was used to reduce the residual "cross-talk" light when the analyzed light was intense and had nearly horizontal polarization plane. The second detector was used when the function of the balanced polarimeter was needed for tests, mathematically simulated with waveforms recorded using a multichannel digital oscilloscope.

The experimental procedure involved recording of waveforms (with sample period of 20 ns) of signals from detectors, representing instantaneous intensities of light of polarizations corresponding to given angular positions of the HWP's polarization axis. One measurement cycle started with the laser beam turned off. The light was suddenly turned on with the AOM for 80 μs. The speed of response of the modulator to the square-wave leading edge of its control signal allowed reaching standby intensities in less than 200 ns. The incident light power was constant during the pulses. The cycles were repeated for 42 positions of the HWP's axis to perform the full linear polarization analysis: extraction of instantaneous total intensities and angular positions of the polarization plane – based on fits of the Malus law formula to the measured signal-vs-angle dependencies.

Data were obtained with the cell temperature of 73.2°C corresponding to the concentration of $2\times10^{11}$ of $^{87}$Rb atoms per cm$^3$ [6]. The laser was tuned to the D1 resonance line (795 nm) and transitions between the ground ($|g>$) $5^2S_{1/2}$ F=2 and the excited ($|e>$) $5^2P_{1/2}$ F'=1 hyperfine structure states, and actively locked at the center of the absorption line. The relevant states are depicted schematically in Fig. 1 (frame) showing also the way a magnetic field splits corresponding degenerate energy levels into the Zeeman sublevels. Levels of the second hyperfine component of the $^{87}$Rb ground state ($|g'>$) $5^2S_{1/2}$ F=1 are also included as important for the optical pumping. Inactive atoms – decoupled from radiation – populated $|g'>$ in result of radiative de-excitation of atoms in $|e>$ (spontaneous emission) or after collisions of atoms in $|g>$ with walls of the cell. The latter was also the way the inactive atoms could become active (|g'>→|g>). The motion of atoms was the only relevant mechanism leading to the relaxation – atoms leaving the irradiated volume depolarized in collisions with walls then returned to the volume with "reset" probability of occupation of the relevant states.

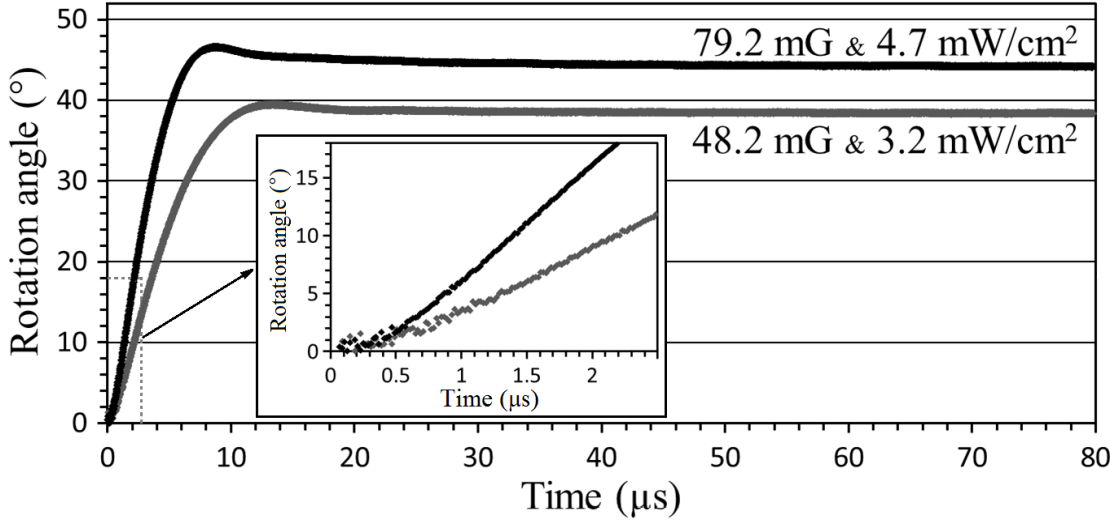

FIG. 2. Polarization plane rotation angle vs time for two example combinations of the magnetic field and the incident laser beam power (estimated intensities at beam's axis are indicated). Standard uncertainties: below 1°, except low-light initial stages (shown in the inset). Data derived from non-averaged waveforms.

The main finding is associated with the fact that rotation of the polarization plane of radiation emerging from atomic vapors appears as a not direct effect of applied magnetic field. Although the rotation angle eventually reaches values characteristic for the strong resonant MOEs, the initial polarization plane is the same as the original (that of the incident beam). The rotation develops with time, in sequences depending on the experimental conditions. Two examples of the time dependence are shown in Fig. 2. Rotation angle equal zero corresponds to the linear polarization of the incident laser beam, time equal zero denotes the beginning of the incoming light pulse. The set of data shown in black is for the magnetic field and the incident beam power selected so the stationary rotation of approximately 45° was attained after about 30 μs. Increasing the magnetic field did not result in any increase of the steady-state rotation. To achieve it, both the magnetic field and the radiation intensity had to be higher. The second set of data (grey symbols) recorded with lower magnetic field and weaker laser beam illustrates corresponding differences in the time dependence.

Observations cannot be explained based on the passive description presuming the rotation angle at a given depth within the medium resulting from combined contributions of preceding layers [4]. The contributions would have to occur in time scales typical for the radiative interactions of atoms – tens of nanoseconds in the considered case ($5^2P_{1/2}$ state lifetime is 27.7 ns [6]). Steady-state rotation angles are large, therefore considerations can be limited to two physical mechanisms associated with the strongest nonlinear MOEs [4]. The Bennet-structure-related ones are not expected to play an important role [7]. Substantial rotation of the polarization plane should be therefore attributed to the internal atomic coherences subject to time evolution under the influence of the magnetic field. Quantum state coherences are induced and/or altered immediately at the moment the radiation is turned on. In the presence of the magnetic field also corresponding modifications of their evolution commences at once. The observed mode of evolution characterized by much lower rate implies involvement of additional mechanisms, certainly of a collective nature. Obviously, it is necessary to include them explicitly in theoretical models focused on time dependencies.

Revolving plane of polarization is not the only time-dependent feature of radiation emerging from the medium. Equally strong dependence, with similar general pace of variations, characterizes the total intensity of light transmitted by vapors. It starts to increase immediately or with some delay (at higher atomic concentrations) when the laser beam is turned on. Fig. 3 (a) shows time dependencies of the power of emerging light derived from the same set of measurements as the rotation angle data above. Corresponding reference data with no applied magnetic field were also plotted for comparison. Full 80-µs irradiation cycle is included to show how the target stationary signal levels were attained. More details of the most dynamic initial stages of the evolution can be seen in Fig. 3 (b). It is clear that even in the field-free conditions, only seemingly simple for description, the change in the input power leads to noticeably distinct way the equilibrium of optical pumping and relaxation is reached.

The conditions of reported measurements are specially interesting because of the fact that the concentration of active atoms allowed observation of both the linear dissipative response of the medium and the diverse routes of the subsequent evolution. Fig. 3 (c) shows the same set of data in the limits of initial 2.5 µs, with common vertical axis for all sets of data (except the scaled dashed line trace depicting the time profile of pulses transmitted with the laser detuned far from the resonance). Approximately 160-ns-long leading edges of all waveforms follow the same path reflecting the speed of AOM's response to its step-function control signal. The same behavior has been confirmed for beams of intensities lower as well as up to 10 times higher than presented. The ratio of heights of the edges and corresponding input beam powers was constant within estimated uncertainties, directly indicating the linearity of initial scattering in the incident power. Time profiles of subsequent evolution of amounts of light emerging from vapors does not quite match what one could expect based on general understanding of optical pumping and relaxation processes involved. With assumed constant number of active atoms arriving to the irradiated volume, the rate of optical pumping should be highest at the beginning and continuously decrease resulting in concave time-profile curves. The ones shown in Fig. 3 (c) are clearly convex initially and turn to concave after a sufficient time which depends on the incident beam intensity. The stronger the input beam the sooner the profile's inflection occurs. It is present also at much higher intensities than ones reported in this work, with and without applied magnetic field.

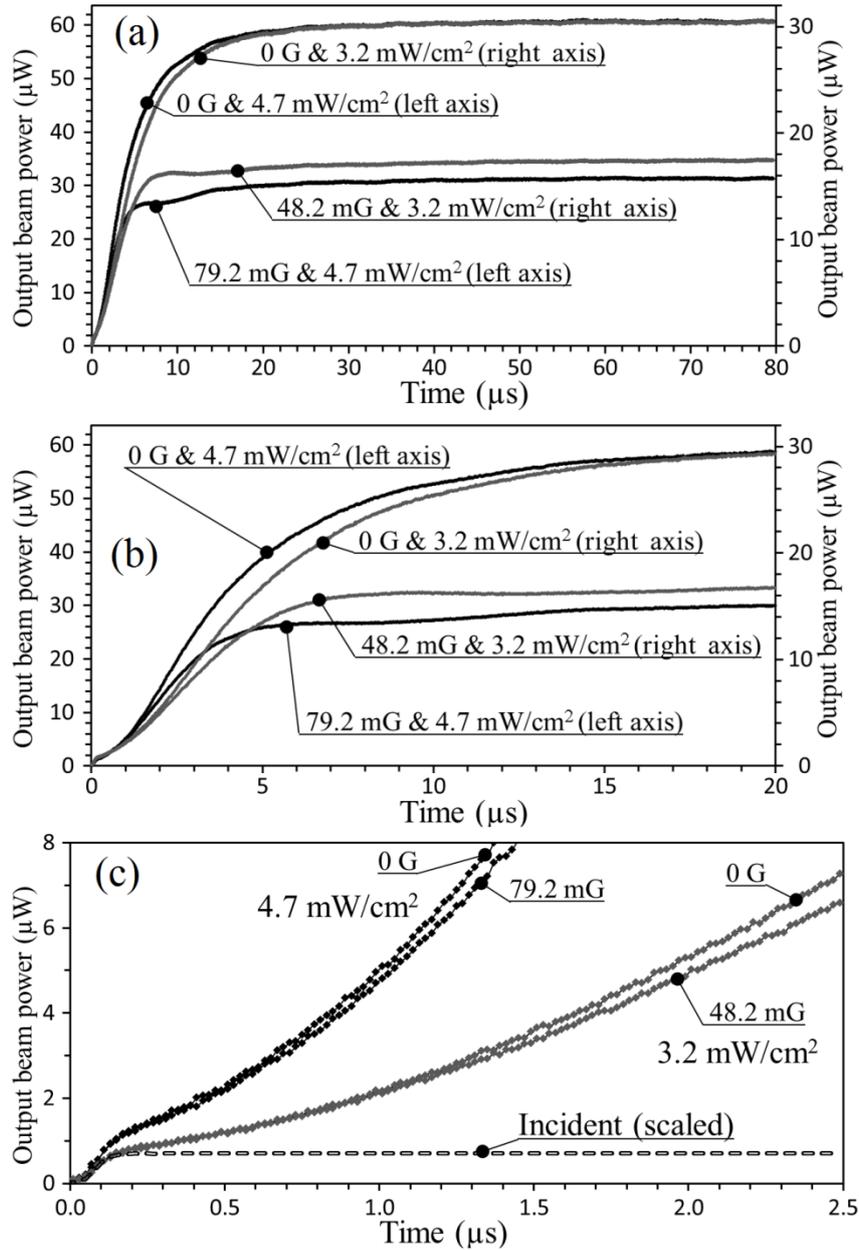

FIG. 3. Total power of beams emerging from atomic vapors vs time for two example combinations of the magnetic field and the incident laser beam power. Corresponding signals with no applied magnetic field are included for comparison. Standard uncertainties below 0.5 µW. Dashed line in (c) – scaled time-profile of incident pulses.

Such the characteristic yet diverse behavior implies that some significant part of the output beam can be attributed to stimulated emission and collective effects. First, an avalanche-like increase is formed (convex parts of profiles) exhausting quickly initially most numerous population of excited atoms. Progressive reduction of the number of atoms the emission can be stimulated on – due to the lasing and the optical pumping – slows down the increase and finally leads to the inflections of profiles. Dynamic equilibria are attained eventually with stationary spatial distributions of light and relevant atomic states populations. The avalanche-like components of the output beams, can be interpreted as forming moving spatial structures closely resembling polaritons from the EIT context [8]. This picture has clearly some direct relation with the forward scattering studied, however, with use of dramatically higher magnetic fields [9]. The concept of the multi-atom coherence have been also used on this background in analysis of properties of media of high optical density [10].

Radiation emerging from the medium has therefore a structure comprised of residual – or stronger at lower atomic concentrations – part of the incident beam and one or more components resulting from lasing induced on the excited atoms distributed in the interaction volume. The ingredients are not distinguishable without a magnetic field acting on atoms. When one is applied, the stimulated emission induces transitions to states which are coherent superpositions of Zeeman sub-states characterized by shifted energies and, importantly, also differently modified phases. This results in rotated polarization plane of the stimulated component of the beam. The lasing leading slope requires time to build up, so the competing, also rising, transmitted beam dominates initially, causes the polarization to be close to the original one (incident). However, the amount of stimulated light increases quickly leading to the observed revolving plane of polarization and final stationary rotation angles of substantial values.

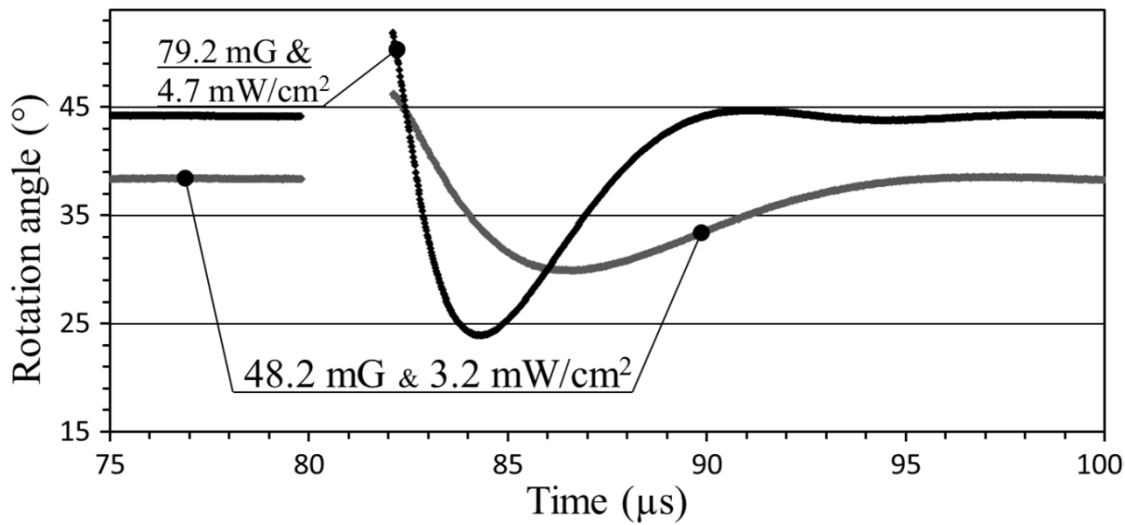

FIG. 4. Example time dependencies of the polarization plane rotation angle just before and after 2-μs intervals with no incident light for two example combinations of the magnetic field and the incident laser beam power.

The competition of transmitted and stimulated components is confirmed by overturns of rotation angles reaching their maxima as can be seen in Fig. 2. Much stronger manifestations can be observed when the laser beam is rapidly turned off and back on after a time which allows only a fraction of atoms in the interaction volume to be exchanged. Fig. 4 illustrates the consequences of 2-μs-long intervals with no incident light, followed by returns of intensities to previous stand-by values. The atom-exchange relaxation ongoing in the darkness affects the transmitted component more than the stimulated one. Due to anisotropy of the process, the central segments in which the stimulation is more likely, lose less coherently prepared atoms than the peripheral ones. New active atoms in the latter absorb most of the light when it re-appears, attenuating the initial transmission. Lasing reconstructs itself almost immediately and dominates the emergent beam, rotating the polarization plane noticeably further than in the stand-by. However, the stimulated component quickly over-exhausts its coherent atomic resources and the increasing transmission takes over again for some time. The subsequent evolution, though via different routes, leads to the previous stand-by spatial distributions of light and the atomic state populations, provided sufficient time is ensured for it.

Selected experimental data on time dependencies of the magnetic polarization plane rotation and the total power of the resonant light beam emerging from suddenly irradiated rubidium vapors were presented. They cannot be explained within the frame of the traditional passive picture of the resonant MOR. The timing of noticeable evolution is a determinant of the incident beam cross section and its intensity profile. Eventually, spatial distributions of relevant atomic states and of the radiation in vapors become stationary. They are mutually dependent and vary in a specific accord before the equilibrium is reached. In realistic situations they are three-dimensional and not trivial. Proper evaluation of their time evolution is the most crucial condition of successful modelling and faithful theoretical reproduction of experimental results. Accounting for it in the analysis of strong resonant interactions with light developing collective atomic coherences concurrently with the optical pumping and simultaneously with the evolution driven by magnetic field is a challenging task due to computational complexity. The problem should be, however, conquerable for the modern sophisticated modelling methods. Experimental data on time-dependent polarization properties of light resulting from coherent interactions with atoms are potentially available in ample amounts with low redundancy. Their theoretical reproduction – complementary to the spectral studies – has the potential to augment the understanding of atomic quantum coherence and possibly facilitate its further applications in science and technology.